\documentclass[fleqn,twoside]{article}
\usepackage{espcrc2,amsmath}

\usepackage{graphicx}
\usepackage[figuresright]{rotating}

\newcommand{\beq}{\begin{equation}}
\newcommand{\eeq}{\end{equation}}
\def \cpp{C_{\pi^+\pi^-}}
\def \spp{S_{\pi^+\pi^-}}
\def\GammaB{\overline{\Gamma}}

\title{Penguin pollution estimates relevant for $\phi_2/\alpha$ extraction}

\author{Jure Zupan \address[IJS]{J.~Stefan Institute, Jamova 39, P.O. Box 3000, 1001
Ljubljana, Slovenia}%
\address[LJU]{Department of Physics, University of Ljubljana, Jadranska 19, 1000
Ljubljana, Slovenia}
\thanks{Based on talks given at "CKM2006", Dec 12th-16th, 2006, Nagoya, Japan, and at
"Beauty 2006", Sept 25th-29th 2006, Oxford, UK.}
}

\begin{document}

\begin{abstract}
A review of methods to extract  the standard CKM unitarity triangle angle $\alpha$ is provided. The sizes of related theoretical errors are reviewed. 
\end{abstract}

\maketitle

\section{Introduction}
The determination of the standard CKM unitarity triangle angle~$\beta=\arg(-V_{cb}^* V_{cd}/V_{tb}^*V_{td})$ from $b\to c\bar cs$ transitions (i.e. $B(t)\to J/\psi K_S$ and related modes) 
has reached an impressive precision \cite{hfag}
\beq
\beta = (21.2\pm1.0)^\circ,
\eeq
with the ultimate theory error 
below percent level \cite{Ciuchini:2005mg}.
This sets high standards for the determinations of the other two angles, $\alpha=\arg(-V_{tb}^*V_{td}/V_{ub}^*V_{ud})$ and $\gamma=\arg(-V_{ub}^*V_{ud}/V_{cb}^* V_{cd})$. In this review we address the theoretical errors for the methods that are used
in the determination of $\alpha$. They represent the ultimate precision at which the angle $\alpha$ can be measured even with unlimited statistics. 

\begin{figure}
\includegraphics[height=3.8cm]{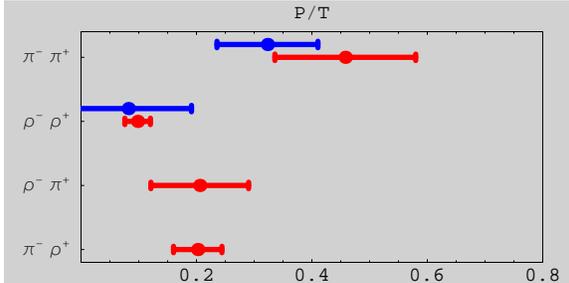}
${}$\\[-17mm]
${}$
\caption{ The sizes of penguin-to-tree ratios $r$ obtained from isospin decomposition (blue) or using SU(3) (red). The very loose bounds on $r({\rho^\pm\pi^\mp})$ following from isospin analysis are not shown.}\label{Fig:r_list}
\end{figure}

The theoretical uncertainties arise because $\sin 2\alpha$ is directly related to measured CP asymmetries only in the limit of one dominant amplitude. Focusing for the moment on $B\to \pi^+\pi^-$ decays for clarity, we define the "tree", $T$, and "penguin", $P$, amplitudes
\beq
A(B^0 \to \pi^+ \pi^-) \equiv A_{\pi^+\pi^-}=  T e^{i \gamma} + P e^{i \delta},\label{TPsplit}
\eeq
and $\bar B^0\to \pi^+\pi^-$ amplitude $\bar A_{\pi^+\pi^-}$ is obtained through $\gamma\to -\gamma$ exchange.
In our notation penguin carries only the strong phase $\delta$, and tree only the weak phase $\gamma$. 
The angle $\alpha$ is extracted  from indirect CP asymmetry 
where
\beq\label{Sf}
\spp= -2 \frac{\Im(\lambda)}{1+|\lambda|^2}, \quad \lambda=e^{-2 i  \beta}\frac{\bar{ A}_{\pi^+\pi^-}}{A_{\pi^+\pi^-}},
\eeq
measured in the
time dependent decay width
\beq
\begin{split}
\Gamma(B^0(t)&\to\pi^+\pi^-) \propto 
{\Gamma_{\pi^+\pi^-}}
\left [1 + \right.\\
+&\left. { \cpp}\cos\Delta mt-{ \spp}\sin\Delta mt\right ], \label{Gamma}
\end{split}
\eeq
Expanding to first order in 
$r=P/T$ one has
\beq
\begin{split}
S_{\pi^+\pi^-}=&\sin 2\alpha +2 r \cos \delta \sin(\beta\negthinspace+\negthinspace\alpha)\cos 2\alpha.
\label{spipi}
\end{split}
\eeq
In the limit of vanishingly small penguins $S_{\pi^+\pi^-}=\sin 2\alpha$, which receives $O(r)$ corrections \eqref{spipi}. The theoretical error on $\alpha$ therefore depends on how well we can determine the hadronic parameters $r$ and $\delta$. It is not possible to determine both of them from $\Gamma(B^0(t)\to\pi^+\pi^-)$. Namely, in \eqref{Gamma} we have 3 measurables that depend on 4 unknowns: $T, r, \delta, \gamma$. Additional input is thus required in order to determine the penguin pollution (i.e. fix the size of $r$). This can be obtained by using approximate symmetries of QCD: isospin, flavor SU(3) or using expansion in $1/m_b$. We describe these three approaches and the resulting errors in the rest of the review. As evident from Fig.  \ref{Fig:r_list} the penguin-to-tree ratios obey a hierarchy
\beq
r(\pi^+\pi^-)>r(\rho^+\pi^-)\sim r(\pi^+\rho^-)>r(\rho^+\rho^-).
\eeq
For methods based on SU(3) and $1/m_b$ we expect the same hierarchy in the theory errors, while this may not be the case for methods based on isospin as will be discussed in more detail below.

${}$\\[-9mm]
${}$

\section{Isospin}
\subsection{Measurement of $\alpha$ from $B\to \pi \pi$}
\begin{figure}[t]
\begin{center}
\includegraphics[width=6.cm]{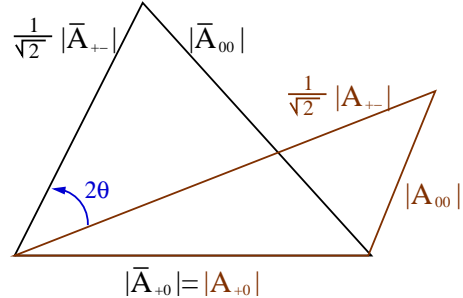}
${}$\\[-17mm]
${}$
\end{center}
\caption{
Gronau-London construction, with only one of four possible triangle orientations shown. The subscripts $A_i$ denote pion charges, while $\bar A_i$ refer to CP conjugated modes, eg. $A_{+-}=\langle \pi^+\pi^-| H| B^0\rangle$.}  \label{ZupanGLtriangle-fig}
\end{figure}

The standard methods for obtaining $\alpha$ from $B\to \pi\pi, \rho\rho, \rho\pi$ all use isospin. The measurement of $\alpha$ from {$B(t)\to \pi\pi$} or longitudinal $\rho\rho$ relies on the triangle construction due to Gronau and London \cite{Gronau:1990ka}, shown on Fig. \ref{ZupanGLtriangle-fig}. The triangles visualize the isospin relations between decay amplitudes. Furthermore, the observable $\sin( 2\alpha_{\rm eff})=S_{\pi\pi}/\sqrt{1-C_{\pi\pi}^2}$
${}$ is directly related to $\alpha$ through
$2\alpha=2 \alpha_{\rm eff}-2 \theta$,
with $\theta$ defined on Fig. \ref{ZupanGLtriangle-fig}. The construction relies on the following assumptions: (i) that $\Delta I=5/2$ operators are not present in the effective weak Lagrangian (these can eg. arise from $\Delta I=2$ 
electromagnetic rescattering of two pions),  (ii) that there are no isospin breaking corrections and (iii) that $A_{+0}$ is a pure tree, so that the bases of the two triangles on Fig. \ref{ZupanGLtriangle-fig} coincide. 

The $\Delta I=5/2$ effects have not been analysed in detail in this context. One expects them to be parametrically small, suppressed by electromagnetic coupling $\alpha_{\rm em}$ and thus $O(1\%)$. The other two corrections are related to isospin breaking due to different $u$ and $d$ quark masses and charges. Isospin breaking has several effects relevant for $\alpha$ determination: 
(i) the basis of operators in effective weak Hamilton is extended to include electroweak penguin operators (EWP) $Q_{7,\dots, 10}$, (ii) $\pi^0$ mass eigenstate no longer coincides with the isospin eigenstate leading to  $\pi^0-\eta-\eta'$ mixing (and similarly to $\rho^0-\omega$ mixing for the vector mesons), (iii) the reduced matrix elements between states in the same isospin multiplet may no longer be related simply by SU(2) Clebsch-Gordan coefficients. Not all of these effects can be bounded at present. In literature only the first two have been analyzed in some detail \cite{Gardner:1998gz,Gardner:2005pq,Gronau:2005pq}. 

The effect of EWP is known quite precisely. Including them still gives two closed triangles as in Fig. \eqref{ZupanGLtriangle-fig}. The triangle bases, however, are now at an angle to each other since $A_{0+}, \bar A_{0+}$ are no longer only pure tree. Neglecting small contributions from $Q_{7,8}$ operators \cite{Gronau:1998fn,Gronau:2005pq}, while relating $Q_{9,10}$ to the tree operators \cite{Gronau:1998fn,Neubert:1998pt},
\beq
\begin{split}
\Delta \alpha&=-\frac{3}{2}\left(\frac{C_9+C_{10}}{C_1+C_2}\right)\frac{\sin(\beta+\alpha)\sin\alpha}{\sin\beta}\\
&=(1.5\pm 0.3\pm{0.3})^\circ,
\end{split}
\eeq
where the first error is from experimental uncertainty on $\alpha, \beta$, while the second is an estimate of the error due to nonzero $Q_{7,8}$ matrix elements. Exactly the same shift due to EWP holds for the $\rho\rho$ system and between $T$ and $\overline T$ tree amplitudes in $B\to \rho\pi$. 

The  effect of $\pi^0-\eta-\eta'$ mixing is known less precisely. Because the $\pi^0$ wave function is composed of $I=1$ and small $I=0$ terms, $|\pi^0\rangle=|\pi_3\rangle +\epsilon |\eta\rangle +\epsilon'|\eta'\rangle$, with $\epsilon = 0.017 \pm 0.003,~\epsilon' = 0.004 \pm 0.001$ \cite{Kroll:2004rs}, there are additional contributions from $A(B^0\to \pi^0\eta^{(')})$ so that the triangles in Fig. \eqref{ZupanGLtriangle-fig} no longer close \cite{Gardner:1998gz}. Varying the phases of these amplitudes, while constraining the magnitudes from experiment leads to a bound \cite{Gronau:2005pq}
\beq
|\Delta\alpha_{\pi-\eta-\eta'}|  < 
1.6^\circ.
\eeq
These two examples of isospin breaking effects show that even though not all of the isospin breaking effects can be calculated or constrained at present, the ones that can be are of expected size
\beq
\sim (m_u-m_d)/\Lambda_{QCD}\sim \alpha_0\sim 1\%.
\eeq

\subsection{Measurement of $\alpha$ from $B\to \rho \rho$}
The isospin analysis in $B\to \rho\rho$ follows the same lines as for $B\to \pi \pi$, but with
three separate isospin relations, one for each polarization. However, in $B\to \rho \rho$ longitudinal component completely dominates so that only
one isospin relation is relevant. Another difference from the $\pi\pi$ system is that $\rho$ resonances have a nonnegligible decay 
width. The two $\rho$ resonances can then also form an $I=1$ final state \cite{Falk:2003uq}, which affects the analysis at 
 $O(\Gamma_\rho^2/m_\rho^2)$, but it is possible to constrain this effect
 experimentally \cite{Falk:2003uq}.

As before the theoretical error on extracted $\alpha$ is given by isospin breaking effects. While the shift due to EWP is exactly the same as in $\pi\pi$, there are also novel effects due to $\rho-\omega$ mixing \cite{Gronau:2005pq}.  Relatively large, $O(1)$ effect is expected near $\omega$ resonance invariant mass of final $\pi^+\pi^-$ pair, while the integrated effect of the $\omega$ resonance is $<2\%$ and is thus of expected size for isospin breaking. There are also other isospin effects, e.g. that couplings between $\rho^+, \rho^0$ and $\pi^+\pi^0, \pi^+\pi^-$ final states may not be given by SU(2) Clebsch-Gordan coefficients. All the effects that can be estimated now, however, are found to be of the expected size of $O(1\%)$ \cite{Gronau:2005pq}

\subsection{$B\to \rho\pi$}
The Snyder-Quinn method exploits the fact that $B\to \rho^\pm \pi^\mp, \rho^0\pi^0$ decays are part of the $B\to\pi^+\pi^-\pi^0$ three body decay \cite{Snyder:1993mx}. From time dependent $B^0$ and $\bar B^0$ Dalitz plot analysis one can therefore measure magnitudes {\it and relative phases} of $B^0 (\bar B^0)\to \rho^\pm \pi^\mp, \rho^0\pi^0$  amplitudes from the overlaps between different $\rho$ resonance bands. This additional information has a beneficial consequence that only an isospin relation between penguin amplitudes is needed (in contrast to Gronau-London triangle constructions in $B\to \pi\pi, \rho\rho$, where also relations between tree amplitudes are used).
Isospin breaking effects are thus expected to be $P/T$ suppressed  \cite{Gronau:2005pq}! (This would not be true for the pentagon analysis, though.) The exception to this rule are EWP that are expected to lead to the largest contribution because they are directly related to the tree amplitudes, but are easily taken into account similarly as in $B\to \pi\pi$. Other isospin breaking effects are expected to be $P/T\sim 0.2$ suppressed. For instance, using SU(3) relations the shift due to $\pi^0-\eta-\eta'$ mixing was estimated in \cite{Gronau:2005pq} to be
\beq
\begin{split}
|\Delta\alpha_{\pi-\eta-\eta'}| &=  \frac{|\epsilon P_{\rho\eta} +  \epsilon' P_{\rho\eta'}|}{|T|} \\
&\le
0.024\epsilon + 0.069\epsilon' \le  0.1^\circ.
\end{split}
\eeq 
This does not include all isospin breaking but it does show that the suppression exists.

\section{SU(3)}
\subsection{$B\to \rho \rho$}
We do not discuss $B\to \pi\pi$ system in this context since penguin pollution is larger, so that SU(3) method is expected to have larger errors. The penguin pollution on the other hand is very small for $B\to \rho\rho$, cf. Fig. \ref{Fig:r_list}. Small penguin pollution makes, somewhat surprisingly, the method based on the SU(3) symmetry as theoretically clean as the isospin analysis \cite{Beneke:2006rb}. The reason is that SU(3) symmetry is used to directly bound $P/T$, while isospin construction involves relations between the tree amplitudes as well. The isospin breaking to this larger amplitudes thus also enter into the corrections. One further advantage of the SU(3) approach over isospin analysis is that one does not need a measurement of $C_{\rho^0\rho^0}$, which is not available yet. (Because of this the isospin analysis at present leads only to a broad constraint $|\alpha-\alpha_{\rm eff}| < 22.4^\circ$ ($95\%$ CL) \cite{tJampens}.)

The basic idea of the method based on SU(3) flavor symmetry is to relate the $\Delta S=0$ $B^0\to \rho^+\rho^-$ decay, in which
tree and penguin amplitudes have CKM elements of similar size $T\sim V_{ub} V_{ud}^*$, $P\sim V_{cb} V_{cd}^*$ to the 
$\Delta S=1$ decays in which the tree is CKM suppressed, $T'\sim V_{ub} V_{us}^*$, while the penguin is CKM enhanced $P'\sim V_{cb} V_{cs}^*$. Thus $P'/T'$ is $1/\lambda^2$ enhanced over $P/T$ and can be used to bound
$P/T$. In Ref. \cite{Beneke:2006rb} the authors used penguin dominated $Br_L(K^{*0}\rho^+)$ for this purpose. 
The relation to the penguin in $B^0\to \rho^+\rho^-$ is then
\beq
|A_L(K^{*0}\rho^+)|^2_{\rm CP-av}= 
F\left(\frac{|V_{cs}|f_{K^*}}{|V_{cd}|f_{\rho}}\right )P^2,
\eeq
where the parameter $F$ is equal to $1$, if SU(3) breaking can be factorized, and if color-suppressed EWP and penguin annihilation contributions are neglected (for $F=1$ one has $P/T=0.10\pm0.02$ showing that penguin pollution is small). Taking a conservative range $0.3 \le F \le 1.5$ and including pre-ICHEP06 exp. errors on $S_L, C_L$ to fit for two unknowns, $\alpha$ and the strong phase $\delta$, the authors of Ref. \cite{Beneke:2006rb} find
$\alpha = [91.2^{+9.1}_{-6.6}~({\rm exp})^{+1.2}_{-3.9}~({\rm th})]^\circ$, where the last error is due to the variation in $F$ and represents the ultimate theoretical error of this approach. It is comparable to the theoretical error in the isospin analysis.

\subsection{$B\to \rho \pi$}
The Snyder-Quinn method uses interference of $\rho$ resonances in $B\to \pi^+\pi^-\pi^0$ to extract the information on relative phase \cite{Snyder:1993mx}. A potential problem is that the $\rho$ resonance bands  do not overlap head-on and one is sensitive to the resonance tails. This can be avoided by using 
just the $\rho^\pm \pi^\mp$ final states and the $SU(3)$ related modes \cite{Gronau:2004tm}.
 Similarly to 
 (\ref{TPsplit}),
the tree and penguin 
contributions are first defined according to their weak phases
\beq
A(B^0\to \rho^\pm\pi^\mp)= e^{i \gamma}T_{\pm}+ 
{P}_{\pm}.
\eeq
In total there are 
8 unknowns: $|T_\pm|$, $|P_\pm|$, $\arg\big(P_\pm/T_\pm\big)$, $\arg\big(T_-/T_+\big)$, $\alpha$, but just 6 observables.
Additional information on penguin contributions can be obtained from SU(3) related
$\Delta S=1$ modes $B^0\to K^{*+}\pi^-, K^+\rho^-$ and $B^+\to K^{*0} \pi^+, K^0 \rho^+$ , in which penguins are CKM enhanced
and tree terms CKM suppressed compared to the $\rho^\pm \pi^\mp$ final state. This gives at $90\%$ CL \cite{Gronau:2004tm}
 \beq
 \begin{split}
&0.16 \le |P_+/T_+| \le 0.24,\\
&0.12 \le |P_-/T_-| \le 0.29,
\end{split}
\eeq
where in order to relate the $\Delta S=1$ and $\Delta S=0$ channels, 
annihilation like topologies were neglected.
Since penguin contributions are relatively
small, the error introduced
 because of the SU(3) breaking on extracted value of $\alpha$ 
are expected be  small, of order $O(\delta_{SU(3)}P_\pm/T_\pm)$. A Monte Carlo study with up to $30\%$ SU(3) breaking on
penguins for instance gives $\sqrt{\langle(\alpha^{\rm out}-\alpha^{\rm in})^2\rangle}\sim 2^\circ$ \cite{Gronau:2004tm}, but further analyses are called for.
Finally, note that unlike the Snyder-Quinn approach the method of using SU(3) symmetry determines $\alpha$ only up to discrete ambiguities. In order to resolve ambiguities an additional assumption had to be used in \cite{Gronau:2004tm}, namely  $\arg(T_-/T_+)<90^\circ$ in agreement with factorization theorems. The improvement on the knowledge of $\alpha$ gained through this constraint was further explored in \cite{Gronau:2004sj}. 

\subsection{$\alpha$ from $a_1^\pm \pi^\mp$}
While isospin analysis is practically impossible for $a_1^\pm \pi^\mp$ since one either has to deal with resonance overlaps in $4$-body final states or with pentagon relations, the extension of the SU(3) method to $a_1(1260)^\pm \pi^\mp $ final states is fairly straightforward and similar to the case of $\rho \pi$ \cite{Gronau:2005kw}. An additional  complication is that the $K_{1A}$ state, which belongs to the same SU(3) multiplet as $a_1$, is an
admixture of two mass eigenstates. As before the method works much better if $P/T$ is smaller. If $P/T\ll 1$ one can use bounds on penguin pollution, otherwise general SU(3) fits are possible (but in this case no suppression of  SU(3) errors is expected). Because of the $K_{1A}$ mixing one needs branching ratios for
at least 3 SU(3) related modes to arrive at bounds, either
$\GammaB(B^0\to K_{1}^{+}(1400)\pi^-)$ and $\GammaB(B^0\to K_{1}^{+}(1200)\pi^-)$, or
$\GammaB(B^+\to K_{1}^{0}(1400)\pi^+)$ and $\GammaB(B^+\to K_{1}^{0}(1200)\pi^+)$, and in addition
$\GammaB(B^0\to a_1^- K^+)$ or $\GammaB(B^+\to a_1^+ K^0)$. At CKM2006 the first measurement of time dependent $B(t)\to a_1^\pm \pi^\mp$ decay width was reported by BaBar \cite{cf. talk by F. Palombo,Aubert:2006gb}.

\section{$1/m_b$}
An interesting way of using $1/m_b$ expansion to obtain precise constraints in the $\bar \rho$, $\bar \eta$ plane was presented in Ref. \cite{Buchalla:2003jr,Buchalla:2004tw}. A crucial observation by Buchalla and Safir was that the penguin, $P$, and tree, $T$, parameters in \eqref{TPsplit} already contain the absolute values of CKM elements $|V_{ub}^*V_{ud}|$ and $|V_{cb}^*V_{cd}|$ respectively. They define a new ratio where the dependence on the CKM elements is explicit
\beq
\frac{P}{T}=r=\frac{\tilde r}{\sqrt{\bar \eta^2+\bar \rho^2}}\simeq\frac{1}{2.5} \tilde r,
\eeq
with $\bar \rho, \bar \eta$ the parameters in a Wolfenstein parametrization of the CKM matrix. One can then get a precise constraint in the $\bar \rho, \bar \eta$ plain, if the hadronic parameters $\tilde r$ and the strong phase $\delta$ are known to some reasonable degree. Changing the variables from $\bar \rho, \bar \eta$ to $\tan\beta, \bar \eta,$ (since $\beta$ was measured precisely from $S_{J/\Psi K_S}$) one gets
\beq
\bar \eta=\frac{1+\cot \beta S-\sqrt{1-S^2}}{(1+\cot^2\beta)S}(1+\tilde r \cos \delta ). \label{etarel}
\eeq
 It is important to note that the unknown hadronic parameter $\tilde r$ enters only in the form $1+ \tilde r \cos\delta $. In QCD factorization its value is $r=0.107\pm0.031$, where the error is the combined error from input parameters and the (model dependent) estimates of uncertainties due to subleading corrections. The error on $\bar \eta$ is thus small, at the order of a few percent.  In the original proposal \cite{Buchalla:2003jr,Buchalla:2004tw} the strong phase $\delta$ was taken to be  small, so that $\cos\delta \to 1$ with corrections only of second order in $\delta$. These are small, if charming penguins are perturbative (in QCD factorization $\delta=0.15\pm.25$ \cite{Buchalla:2004tw}). Alternatively one could determine the strong phase $\delta$ from $C_{\pi^+\pi^-}$ simultaneously with $\bar \eta$ in \eqref{etarel}. $1/m_b$ expansion was used also in Ref. \cite{Bauer:2004dg} in a different way, as an additional constrain on the isospin analysis, while in Ref. \cite{Datta:2006af} it was argued that $1/m_b$ expansion can be used to extract $\alpha$ from $B\to K^0\bar K^0$, albeit with relatively large theoretical errors.


\begin{thebibliography}{99}
\bibitem{hfag}
Summer 2006 HFAG average, available at www.slac.stanford.edu/xorg/hfag/

\bibitem{Ciuchini:2005mg}
  M.~Ciuchini, M.~Pierini and L.~Silvestrini,
  Phys.\ Rev.\ Lett.\  {\bf 95}, 221804 (2005)
  [arXiv:hep-ph/0507290];
  H.~Boos, T.~Mannel and J.~Reuter,
  Phys.\ Rev.\ D {\bf 70}, 036006 (2004)
  [arXiv:hep-ph/0403085];
  H.~n.~Li and S.~Mishima,
  arXiv:hep-ph/0610120;
  Y.~Grossman, A.~L.~Kagan and Z.~Ligeti,
  Phys.\ Lett.\ B {\bf 538}, 327 (2002)
  [arXiv:hep-ph/0204212].
  
\bibitem{Gronau:1990ka}
M.~Gronau and D.~London,
Phys.\ Rev.\ Lett.\  {\bf 65}, 3381 (1990).
  
\bibitem{Gardner:1998gz}
S.~Gardner,
Phys.\ Rev.\ D {\bf 59}, 077502 (1999).

\bibitem{Gronau:2005pq}
  M.~Gronau and J.~Zupan,
  Phys.\ Rev.\ D {\bf 71}, 074017 (2005)
  [arXiv:hep-ph/0502139].
  
  
\bibitem{Gardner:2005pq}
  S.~Gardner,
  Phys.\ Rev.\ D {\bf 72}, 034015 (2005)
  [arXiv:hep-ph/0505071].

\bibitem{Gronau:1998fn}
M.~Gronau, D.~Pirjol and T.~M.~Yan,
Phys.\ Rev.\ D {\bf 60}, 034021 (1999)
[Erratum-ibid.\ D {\bf 69}, 119901 (2004)].



\bibitem{Neubert:1998pt}
M.~Neubert and J.~L.~Rosner,
Phys.\ Lett.\ B {\bf 441}, 403 (1998);
A.~J.~Buras and R.~Fleischer,
Eur.\ Phys.\ J.\ C {\bf 11}, 93 (1999).

\bibitem{Kroll:2004rs}
  P.~Kroll,
  Int.\ J.\ Mod.\ Phys.\ A {\bf 20}, 331 (2005)
  [arXiv:hep-ph/0409141].
  
\bibitem{Falk:2003uq}
A.~F.~Falk, Z.~Ligeti, Y.~Nir and H.~Quinn,
Phys.\ Rev.\ D {\bf 69}, 011502 (2004).

\bibitem{Snyder:1993mx}
A.~E.~Snyder and H.~R.~Quinn,
Phys.\ Rev.\ D {\bf 48}, 2139 (1993).

\bibitem{Beneke:2006rb}
  M.~Beneke, M.~Gronau, J.~Rohrer and M.~Spranger,
  Phys.\ Lett.\ B {\bf 638}, 68 (2006)
  [arXiv:hep-ph/0604005].
  
\bibitem{tJampens}
S. T'Jampens, talk at "Beauty 2006",
Sept 25th-29th 2006, 
Oxford, UK.

\bibitem{Gronau:2004tm}
  M.~Gronau and J.~Zupan,
  Phys.\ Rev.\ D {\bf 70}, 074031 (2004)
  [arXiv:hep-ph/0407002].
  
\bibitem{Gronau:2004sj}
  M.~Gronau, E.~Lunghi and D.~Wyler,
  Phys.\ Lett.\ B {\bf 606}, 95 (2005)
  [arXiv:hep-ph/0410170].
  
\bibitem{Gronau:2005kw}
  M.~Gronau and J.~Zupan,
  Phys.\ Rev.\ D {\bf 73}, 057502 (2006)
  [arXiv:hep-ph/0512148].
  
\bibitem{cf. talk by F. Palombo}
F. Palombo, talk at "CKM2006", Dec 12th-16th, 2006, Nagoya, Japan.

\bibitem{Aubert:2006gb}
  B.~Aubert  [BABAR Collaboration],
  arXiv:hep-ex/0612050.


  
\bibitem{Buchalla:2003jr}
  G.~Buchalla and A.~S.~Safir,
  Phys.\ Rev.\ Lett.\  {\bf 93}, 021801 (2004)
  [arXiv:hep-ph/0310218].
  
\bibitem{Buchalla:2004tw}
  G.~Buchalla and A.~S.~Safir,
  Eur.\ Phys.\ J.\ C {\bf 45}, 109 (2006)
  [arXiv:hep-ph/0406016].
  
\bibitem{Bauer:2004dg}
  C.~W.~Bauer, I.~Z.~Rothstein and I.~W.~Stewart,
  Phys.\ Rev.\ Lett.\  {\bf 94}, 231802 (2005)
  [arXiv:hep-ph/0412120].

\bibitem{Datta:2006af}
  A.~Datta, M.~Imbeault, D.~London and J.~Matias,
  Phys.\ Rev.\  D {\bf 75}, 093004 (2007)
  [arXiv:hep-ph/0611280].




\end{thebibliography}
\end{document}